\documentclass[a4paper,twocolumn,superscriptaddress,amsmath]{revtex4-1}
\usepackage{soul}

\usepackage{oubraces}
\usepackage{adjustbox}
\usepackage{color}
\usepackage{graphicx}
\usepackage{subfigure}
\usepackage{amsmath}
\usepackage{amsfonts}
\usepackage{citesort}
\usepackage{bm}
\usepackage{mathrsfs}

\newcommand{\comment}[1]{}

\newcommand{\eg}{{e.\,g.\ }}
\newcommand{\ie}{{i.\,e.\ }}

\newcommand{\FreeEnthalpy}{{\mathscr{\tilde G}}}

\newcommand{\Energy}{\mathscr{E}}
\newcommand{\Fermi}{\mathrm{F}}

\newcommand{\sech}{\mathrm{sech}}
\newcommand{\eps}{\varepsilon}
\newcommand{\imag}{\mathrm{i}}
\newcommand{\grad}{\mathrm{grad}}
\newcommand{\total}{\mathrm{d}}

\newcommand{\sgn}{\mathrm{sgn\,}}

\begin{document}

\title{
Enhanced ponderomotive force in graphene due to interband resonance
}
%\title{Stimulated polariton scattering: an avenue to the optical spectroscopy of unhybridized THz-polaritons }

\author{C.~Wolff }
\email{cwo@mci.sdu.dk}
\affiliation{Center for Nano Optics, University of Southern Denmark, 
Campusvej 55, DK-5230 Odense M, Denmark}

\author{C.~Tserkezis }
\affiliation{Center for Nano Optics, University of Southern Denmark, 
Campusvej 55, DK-5230 Odense M, Denmark}

\author{N.~Asger Mortensen }
\affiliation{Center for Nano Optics, University of Southern Denmark, 
Campusvej 55, DK-5230 Odense M, Denmark}
\affiliation{Danish Institute for Advanced Study, University of Southern 
Denmark, Campusvej 55, DK-5230 Odense M, Denmark}
\affiliation{Center for Nanostructured Graphene, Technical University of Denmark, DK-2800 Kongens~Lyngby, Denmark}

\date{\today}

\begin{abstract}
We analyze intrinsic nonlinearities in two-dimensional polaritonic materials 
interacting with an optical wave. 
Focusing on the case of graphene, we show that the second-order nonlinear
optical conductivity due to carrier density fluctuations associated with the 
excitation of a plasmon polariton is closely related to the ponderomotive 
force due to the oscillating optical field.
This relation is first established through an elegant thermodynamic approach 
for a Drude-like plasma, in the frequency range where intraband scattering is 
the dominant contribution to conductivity.
Subsequently, we extend our analysis to the interband regime, and show that for 
energies approximately half the Fermi energy, the intraband contribution to 
the ponderomotive force diverges. 
In practice, thermal broadening regularizes this divergence as one would expect,
but even at room temperature typically leaves a strong ponderomotive 
enhancement.
Finally, we study the impact of nonlocal corrections and find that nonlocality does not
lead to further broadening (as one would expect in the case of Landau damping),
but rather to a splitting of the ponderomotive interband resonance.
Our analysis should prove useful to the open quest for exploiting 
nonlinearities in graphene and other two-dimensional polaritonic materials,
through effects such as second harmonic generation and photon drag.
\end{abstract}

\maketitle

%%%%%%%%%%%%%%%%%%%%%%%%%%%%%%%%%%%%%%%%%%%%%%%%%%%%%%%%%%%%%%%%%%%%%%%%%%%%%%%%
\section{Introduction} 

The emergence of graphene~\cite{CastroNeto:2009,Ferrari:2015} and other two-dimensional (2D)
materials~\cite{Xu:2013,Butler:2013} at the forefront of research in all areas of condensed matter
physics, owing to their intriguing mechanical~\cite{Frank:2007}, thermal~\cite{Balandin:2011},
electronic~\cite{Fiori:2014} and optical properties~\cite{Xia:2014}, has led to a plethora of
suggested applications, many of which are already starting to see the light of day. In photonics,
in particular, where the possibility to excite and tailor highly confined polaritons can have important
implications~\cite{Mills:1974,Zayats:2003}, 2D materials became very quickly prominent templates for
enabling and tailoring light-matter interactions~\cite{Goncalves:2016,Xiao:2016,Low:2017}. In this context, the prospect
of enhanced nonlinearities is always among the first effects to be explored in a novel architecture, and
2D materials could not fail to attract their share of attention~\cite{Hendry:2010,Gullans:2013,Echarri:2018,Cheng:2018,You:2019}.

In the long list of nonlinear optical effects, such as higher-harmonic generation, stimulated Raman
and Rayleigh scattering, electrooptic effects and multiphoton absorption~\cite{Boyd2003}, ponderomotive
effects are particularly relevant to plasmonics. Charged particles in inhomogeneous oscillating electromagnetic
fields are known to be subject to a ponderomotive force proportional to the gradient of the electric field
($ \vec F\propto \grad |\vec E|^2$) that accelerates them towards the field direction~\cite{Landau:1984}. This
has been exploited, for instance, for electron acceleration with laser pulses~\cite{Malka:1996}, and controlling
excitons~\cite{Leinss:2008} or plasmons~\cite{Irvine:2006}, as well as inducing nonlinear effects in
them~\cite{Johnsen:1999,Ginzburg:2010}. With the advent of graphene as an exemplary plasmonic medium, it was only
natural to explore its capability to enhance nonlinearities~\cite{Mikhailov:2011,Hong:2013,Cheng:2014}, and the
ponderomotive force can be a mechanism to lead to, otherwise symmetry-prohibited, second harmonic generation~(SHG)~\cite{Gullans:2013}.

Recently, the second-order nonlinear ac conductivity of a generic Dirac fluid was connected
to the ponderomotive force through the hydrodynamic equations of motion~\cite{Sun:Pnas2018}.
As a directly available test bed, the authors of Ref.~\onlinecite{Sun:Pnas2018} applied their
analysis to graphene, focusing on SHG and photon drag. While they limited themselves to the energy
regime dominated by intraband transitions, they anticipated that at high (e.g. room) temperatures the
interband contribution to the conductivity could become important. This is exactly the focus of this
paper. We start by deriving the connection between the ac conductivity and the ponderomotive force with
a different starting point, through a general and powerful thermodynamic approach which, to the best of
our knowledge, has not been presented before. By introducing the second-order, room-temperature expression
for the conductivity of graphene, we show that the resulting ponderomotive force exhibits a resonance at
an energy twice the Fermi energy, becoming infinite at zero temperature. This resonant behavior survives
for higher temperatures, and can lead to forces as large as one order of magnitude stronger than in the
intraband case, over a relatively wide energy range.

%%%%%%%%%%%%%%%%%%%%%%%%%%%%%%%%%%%%%%%%%%%%%%%%%%%%%%%%%%%%%%%%%%%%%%%%%%%%%%%%
\section{Preliminaries} 

We consider a graphene monolayer sandwiched between two dielectrics with relative permittivities $\eps_1$
and $\eps_2$. Within a local response approximation, the electromagnetic properties of graphene are
characterized by a complex sheet conductance $\sigma(\omega)$. 
It has two contributions ($\sigma = \sigma_{1} + \sigma_{2}$): firstly a Drude model
\begin{subequations}
\begin{align}
\frac{ \sigma_{1} (\omega)}{\sigma_{K}} = & 
\frac{\imag 2 \Energy_\Fermi}{\hbar (\omega + \imag \gamma)},
\label{eqn:sigma1_def}
\end{align}
due to intraband scattering of free carriers. 
It is characterized by a Ohmic damping constant $\gamma$ due to the scattering of
carriers predominantly off lattice impurities at low temperatures and phonons at higher
temperatures.
The second contribution describes the effect of interband transitions if the photon
energy $\hbar\omega$ exceeds twice the Fermi energy $\Energy_\Fermi$ (relative 
to the undoped state). 
At zero temperature. it takes the form:
\begin{align}
\frac{\sigma_{2}^{(T=0)} (\omega)}{\sigma_{K}} = &
\frac{\pi}{2} \left(\Theta(\hbar \omega - 2\Energy_\Fermi) +
\frac{\imag}{\pi} \ln \left| 
\frac{\hbar \omega - 2 \Energy_\Fermi}{\hbar \omega + 2 \Energy_\Fermi}
\right| \right) .
\label{eqn:sigma2_def}
\end{align}
\end{subequations}
Here, $\hbar$ is Planck's constant, $\Theta(x)$ is the Heaviside function, and
$\sigma_{K} = e^{2}/h \simeq 0.387 \times 10^{-5}$\,S is the inverse of the
von-Klitzing constant, with $e$ being the electron charge~\cite{Goncalves:2016}.

The full optical conductivity of graphene at finite temperature has been calculated in
Refs.~\onlinecite{Stauber:2008,Chang:2016}
\begin{widetext}
\begin{align}\label{eqn:sigma_roomT}
\frac{\sigma^{(T>0)} (\omega)}{\sigma_{K}} = & \frac{\imag  2\Energy_\Fermi} { \hbar(\omega + \imag \gamma)} 
+
\frac{\pi}{4} \Bigg(\tanh \frac{\hbar \omega + 2 \Energy_\Fermi}{4 k_{\mathrm{B}} T} +
\tanh \frac{\hbar \omega - 2 \Energy_\Fermi}{4 k_{\mathrm{B}} T} 
%-\frac{\imag}{\pi} \ln\frac{\left(\hbar \omega + 2 \Energy_\Fermi \right)^{2}}{\left(\hbar \omega - 2 \Energy_\Fermi \right)^{2} + \left(2 k_{\mathrm{B}}T \right)^{2}}
+
\frac{\imag}{\pi} \ln\frac
{\left(\hbar \omega - 2 \Energy_\Fermi \right)^{2} + \left(2 k_{\mathrm{B}}T \right)^{2}}{\left(\hbar \omega + 2 \Energy_\Fermi \right)^{2}}
\Bigg),
\end{align}
\end{widetext}
where $k_{\mathrm{B}}$ is the Boltzmann constant and $T$ the temperature. This expression can be further
corrected through appropriate multiplicative factors proportional to the square of the energy over the
hopping parameter of the tight-binding description of graphene~\cite{Stauber:2008}. However, this
correction is usually of minor importance~\cite{Mikhailov:2011,Chang:2016}, and our calculations showed
that it can be safely disregarded here as well.

Graphene is best known and has received most of its attention because of its linear energy-momentum relation
at the undoped Fermi energy~\cite{Goncalves:2016}. This implies that the common expression $m_\mathrm{eff} = 
\hbar^{2} \left[\partial^{2} \Energy / \partial k^{2} \right]^{-1}$ for the electron effective mass is ill-defined
for its quasiparticles, which in fact have vanishing ``rest mass''. Instead, the appropriate expression for
the effective mass is the dynamic mass of a relativistic massless particle, where the energy-independent Fermi
velocity $v_\Fermi$ takes the role of the speed of light~\cite{CastroNeto:2009}:
\begin{align}
\Energy = m_\mathrm{eff} v_\Fermi^{2}.
\label{eqn:emc2}
\end{align}
In further analogy to relativistic massless particles, the Fermi wave number $q_\Fermi$ is related to
the Fermi energy via the linear relationship
\begin{align}
\Energy_\Fermi = \hbar v_\Fermi q_\Fermi.
\label{eqn:Ef_kf_def}
\end{align}
Finally, the Fermi wave number is directly connected to the density $n$ of free carriers:
\begin{align}
q_\Fermi = \sqrt{\pi n}.
\label{eqn:kf_n_def}
\end{align}

Here, we study how a plasmon polariton and an optical wave interact in a graphene sheet due to the intrinsic
nonlinearity of graphene. To this end, we introduce a separation of scales. The optical wave is assumed to
oscillate at an angular frequency $\omega$, whereas the polariton oscillates at an angular frequency
$\Omega \ll \omega$. As a result, the optical parameters can be assumed to be modified by the
polariton, but remain quasi-stationary as far as the optical wave is concerned. 
This separation of frequency scales is intrinsic to the notion of the ponderomotive force. 
Finally, our thermodynamic argument assumes a
macrocanonic ensemble characterized by the free enthalpy $\FreeEnthalpy$.

%%%%%%%%%%%%%%%%%%%%%%%%%%%%%%%%%%%%%%%%%%%%%%%%%%%%%%%%%%%%%%%%%%%%%%%%%%%%%%%%
\section{Intraband contribution to nonlinear response}

First, we derive the ponderomotive force ignoring interband transitions.
The basic idea is that a plasmon polariton is in essence a fluctuation in the 
carrier density, and therefore accompanied by a spatial modulation of the 
local Fermi level $\Energy_\Fermi$.
The notion of a local Fermi level is acceptable as long as the polaritonic 
wavelength is much larger than the lattice constant of the graphene molecule, 
\ie whenever local response theory applies.
It also implies the assumption that the polariton does not cause too much
``unrest'' in the electron system, but rather moves carriers around in a 
quasi-adiabatic way while overall maintaining the general shape of the Fermi 
distribution.
This can be expected whenever the polariton experiences low loss, as any 
qualitative distortion of the Fermi distribution really means dissipation.
The Fermi level in turn controls the intraband conductance of graphene through
Eq.~\eqref{eqn:sigma1_def} and is linked to the total carrier density through
Eq.~\eqref{eqn:kf_n_def}.
This means that we can estimate the change to the optical response simply 
through the chain rule:
\begin{align}
  \frac{\partial \sigma_1}{\partial n} = &
 \frac{\partial \sigma_1}{\partial \Energy_\Fermi }
 \frac{\partial \Energy_\Fermi}{\partial n} 
 = 
 \frac{\imag e^2 v_\Fermi}{2 \hbar (\omega + \imag \gamma) \sqrt{\pi n}}
 = \frac{\sigma_1}{2n}.
 \label{eqn:nonlinearity}
\end{align}

We now aim to gain a better understanding of the expression 
Eq.~\eqref{eqn:nonlinearity} for the polariton-induced change to the optical 
properties.
To this end, we recall the definition of a permittivity from the total free
enthalpy $\FreeEnthalpy$ of a solid~\cite{Landau:1984}:
\begin{align}
  [\eps_r]_{ij} = -\frac{1}{\eps_0} \cdot 
  \frac{\partial^2 \FreeEnthalpy}{(\partial E_i) (\partial E_j)}.
  \label{eqn:eps_def}
\end{align}
Using the relationship $\sigma = -\imag\omega \eps_0 \eps_r$ between the
complex conductance and the permittivity of a material, we can adapt 
Eq.~\eqref{eqn:eps_def} to the case of an isotropic sheet conductance:
\begin{align}
  \sigma = 
  \imag \omega \frac{\partial^2 \FreeEnthalpy}{\partial |\vec E_\parallel|^2},
  \label{eqn:sigma_def}
\end{align}
where $\vec E_\parallel$ is the tangential field at the position of the sheet.
The fact that $\vec E_\parallel$ is continuous across the sheet motivates why
$\vec E$ was chosen as the independent variable~\footnote{
  Strictly speaking, our thermodynamic potential $\FreeEnthalpy$ is not the
  free enthalpy, which is defined with the electric induction $\vec D$ as the 
  independent variable and connected to $\FreeEnthalpy$ via a Legendre
  transformation~\cite{Landau:1984}.
  This detail is of no concern for our argument.
}
in Eq.~\eqref{eqn:eps_def}.

Using Eq.~\eqref{eqn:sigma_def}, we can characterize the intrinsic
second-order nonlinearity of graphene by a parameter $W$ via:
\begin{align}
  W = & 
  \frac{\partial}{\partial n} 
    \left[ \frac{\partial^2 \FreeEnthalpy}{\partial |\vec E_\parallel|^2} \right]
  = \frac{\sigma_1}{2\imag n \omega}.
  \label{eqn:W_parameter}
\end{align}
Under the assumption that this nonlinear process in itself is reversible,
we can interchange the order of derivatives and also find
\begin{align}
  W = & \frac{\partial^2}{\partial |\vec E_\parallel|^2} 
  \left[ \frac{\partial \FreeEnthalpy}{\partial n} \right].
  \label{eqn:ponderomotive_first}
\end{align}
This time, the term in brackets is the definition of a local chemical potential,
so Eq.~\eqref{eqn:ponderomotive_first} describes a correction to the chemical 
potential caused by a change in the intensity of a quickly oscillating 
electric field.
This is of course just a Maxwell relation and intimately related to the
Manley--Rowe relations known from nonlinear optics and electrical engineering \cite{Boyd2003}.
Even though it sounds similar to our starting point (change of the Fermi level 
due to a propagating polariton), it describes in fact the inverse process.
Again, it should be stressed that this conclusion requires the nonlinear
interaction to be reversible, \ie not to create any entropy, not to be 
dissipative.
Therefore, we can expect this to hold exactly in the limit $\gamma = 0$ and
gradually be broken as $\gamma$ assumes a finite value.

From Eq.~\eqref{eqn:nonlinearity}, we can see that $W$ does not depend on the
electric field, so we can derive from Eq.~\eqref{eqn:ponderomotive_first} an
explicit expression for the change in chemical potential
\begin{align}
  \Delta \mu = & \frac{\partial \FreeEnthalpy}{\partial n} 
  = W |\vec E_\parallel|^2.
\end{align}
This exerts a force on each particle that is proportional to the in-plane
gradient:
\begin{align}
  \vec F_1 = & \grad_\parallel (\Delta \mu) 
  = 
 \frac{\sigma_1}{2\imag n \omega} \grad_\parallel
  |\vec E_\parallel|^2,
\end{align}
where $\grad_\parallel$ represents the 2D gradient operator in the sheet plane.
Finally, in the limit of vanishing loss ($\gamma \rightarrow 0$) and using
Eqs.~(\ref{eqn:emc2}--\ref{eqn:kf_n_def}), we find:
\begin{align}
  \vec F = & \frac{e^2}{2 m_\text{eff} \omega^2} 
  \grad_\parallel |\vec E_\parallel|^2.
\end{align}
This is the expression for the ponderomotive force in a 2D plasma composed of 
particles with effective mass $m_\text{eff}$.
Therefore, we have established that the intrinsic nonlinear change of the 
optical conductance due to a plasmon polariton in a graphene sheet is the 
complementary process to the ponderomotive force through which variations in 
the intensity of an optical field can drive polaritons.

A similar expression was recently derived in Ref.~\onlinecite{Sun:Pnas2018}, albeit with a different starting
point and analysis, and restricted to the intraband scattering region. Nevertheless, the authors
of Ref.~\onlinecite{Sun:Pnas2018} did stress the necessity to explore interband corrections, already proven
to be capable of enhancing third-order nonlinearities~\cite{Peres:2014} or contributing to difference
frequency generation~\cite{Yao:2014}, and this is what we shall do in the next section.

%%%%%%%%%%%%%%%%%%%%%%%%%%%%%%%%%%%%%%%%%%%%%%%%%%%%%%%%%%%%%%%%%%%%%%%%%%%%%%%%
\section{Interband contribution}

\begin{figure}
\includegraphics[width=\columnwidth]{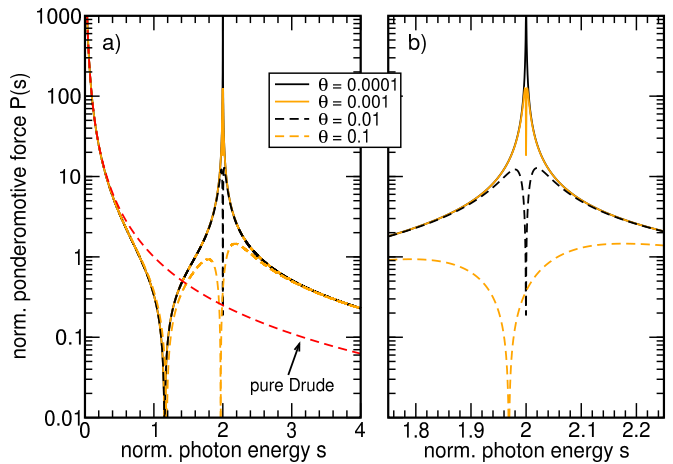}
\caption{
Dependence (overview in panel a, close-up of the interband resonance in panel 
b) of the frequency-part $P(s)$ of the finite-temperature ponderomotive force 
as defined in Eq.~\eqref{eqn:normalized_ponderomotive} on the normalized photon energy
$s = \hbar \omega / \Energy_\Fermi$ for the annotated values of the normalized 
temperature $\theta = k_\mathrm{B} T / \Energy_\Fermi$.
For $\Energy_\Fermi = 893\,\text{meV}$, the temperatures correspond to 
$1, 10, 100,$ and $1000\,\text{K}$, respectively.
The red dashed line indicates the pure Drude-like ponderomotive force without
interband contributions for comparison.
}
\label{fig:ponderomotive}
\end{figure}

The connection we established in the previous section now allows us to find
the generalized ponderomotive force in situations where the electron system
can no longer be described as a Drude plasma.
This is for example the case in the regime $2 \hbar \omega \gtrsim \Energy_\Fermi$, where
the electromagnetic response is significantly modified by interband 
transitions.
Following our previous analysis, we can express the
correction to the ponderomotive force in terms of the density derivative of
the interband conductance, provided this nonlinear coefficient describes a
reversible process.
For optical frequencies $\omega \neq 2 \Energy_\Fermi / \hbar$, we find:
\begin{subequations}
\begin{align}
  \vec F_2^{(T=0)} = & \frac{1}{\imag \omega} \cdot 
  \frac{\partial \sigma_2^{(T=0)}}{\partial n} 
  \grad_\parallel |\vec E_\parallel|^2
  \label{eqn:sigma2_ponderomotive_basic}
  \\
  = &
  \frac{\hbar e^2 v_\Fermi^2}{8 \Energy_\Fermi^2 - 2 (\hbar \omega)^2}
  \ \grad_\parallel |\vec E_\parallel|^2
  .
  \label{eqn:sigma2_ponderomotive_zeroT}
\end{align}

\end{subequations}
This expression is purely real-valued, \ie the variation in the Fermi level
does not lead to a change in the absorptivity of the material unless the
optical frequency is chosen such that $\hbar \omega - 2 \Energy_\Fermi$ changes sign.
Only in this case, the nonlinearity features a substantial imaginary part, \ie 
becomes dissipative.
Otherwise, Eq.~\eqref{eqn:sigma2_ponderomotive_zeroT} constitutes a real ponderomotive 
force.
Due to the resonant nature of Eq.~\eqref{eqn:sigma2_ponderomotive_zeroT}, this force
diverges as the optical frequency approaches the interband threshold from
either side, especially from the low-loss side.
In theory, the ponderomotive forces per unit of optical intensity can be made 
arbitrarily large by moving towards the point $\omega = 2 \Energy_\Fermi /\hbar$. 
This divergence is of course just an artifact of assuming the step-like Fermi 
distribution at zero temperature and will be regularized for any $T \neq 0$.

Indeed, starting from Eq.~\eqref{eqn:sigma_roomT}, we find that the real pole turns
into a Lorentzian imaginary part:
\begin{widetext}
\begin{align}
\frac{1}{\sigma_K}\frac{\partial \sigma_2^{(T>0)}}{\partial \Energy_\Fermi}
= & \frac{\pi }{8 k_\mathrm{B} T} \left[ 
\sech^2 \frac{\hbar \omega + 2 \Energy_\Fermi}{4 k_\mathrm{B} T}
- \sech^2 \frac{\hbar \omega - 2 \Energy_\Fermi}{4 k_\mathrm{B} T}
\right] 
+ \frac{\imag }{\hbar \omega + 2 \Energy_\Fermi} 
+ \frac{\imag(\hbar \omega - 2 \Energy_\Fermi)}{(\hbar \omega - 2 \Energy_\Fermi)^2 + (2 k_\mathrm{B} T)^2}.
\label{eqn:sigma2_ponderomotive_roomT}
\end{align}
Due to the more broadband loss of interband transitions at finite temperature, 
this derivative is non-real everywhere, so strictly speaking, it does not 
provide a real ponderomotive force.
However, at least in parameter ranges where the real part of 
Eq.~(\ref{eqn:sigma2_ponderomotive_roomT}) is small, we can regard inserting 
its imaginary part in Eq.~(\ref{eqn:sigma2_ponderomotive_basic}) to be a good 
approximation:
\begin{align}
\vec F_{2}^{(T>0)}  \approx &
\frac{\hbar e^2}{4 \omega m_\text{eff}}
\Bigg[ 
\frac{\hbar \omega - 2 \Energy_\Fermi}{(\hbar \omega - 2 \Energy_\Fermi)^2 + (2 k_\mathrm{B} T)^2}
+ \frac{1}{\hbar \omega + 2 \Energy_\Fermi}
\Bigg]
\ \grad_\parallel |\vec E_\parallel|^2
  .
\end{align}
\end{widetext}
It is natural to relate both the photon energy $\hbar \omega$ and temperature $T$ to
the Fermi energy in order to obtain a more universal expression.
Introducing the normalized quantities $s = \hbar \omega / \Energy_\Fermi$ and 
$\theta = k_\mathrm{B} T / \Energy_\Fermi$, we find for the total ponderomotive force:
\begin{align}
\vec F_\text{tot}^{(T>0)} = & \overunderbraces{&\br{2}{=\vec F_1(s)}&}
{&\frac{\hbar^2 e^2}{2 m_\text{eff} \Energy_\Fermi^2} 
\grad_\parallel |\vec E_\parallel|^2 \cdot & \frac{1}{s^2} & \cdot 
\left[ 1 + \frac{s(s - 2)}{(s-2)^2 + 4 \theta^2} + \frac{s}{s + 2} \right]
}
{&&\br{2}{= P(s)}}
,
\label{eqn:normalized_ponderomotive}
\end{align}
where the factor outside the bracket turns out to be the Drude-like intraband
ponderomotive force $\vec F_1$ as indicated.
The dimensionless function $P(s)$ summarizes the frequency-dependence and describes the ponderomotive
force normalized to all appearing fundamental constants and the electric field gradient.
In Fig.~\ref{fig:ponderomotive}, we show $P(s)$ for a number of normalized 
temperatures $\theta$.

\section{Nonlocal corrections}

\begin{figure}
  \includegraphics[width=\columnwidth]{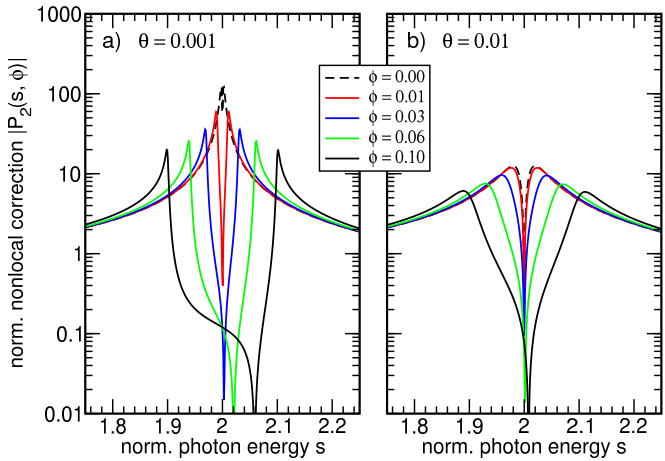}
  \caption{
  Effect of nonlocality on the interband contribution to the ponderomotive force [Eq.\eqref{eqn:nonlocal_approx}] at two normalized temperatures (panel a: $\theta=0.001$; panel b: $\theta=0.01$) and
  for several values of $k_\parallel$ corresponding to different angles of incidence $\phi=k_\parallel/k_\Fermi$ of the driving light field.
  }
  \label{fig:nonlocal_approx}
\end{figure}

For completeness, we finally present the main effect of nonlocality to the ponderomotive
force at finite temperature.
Since the ponderomotive force describes a nonlinearity, terms like ``nonlocality'' or
``inhomogeneity'' require clarification to avoid confusion.
We consider the dependence of the ponderomotive force on the \emph{optical} wave vector
$\vec k_\parallel$ projected to the sheet plane.
In analogy to the scale separation in time mentioned in the preliminaries, we assume
that the length scale of the charge distribution (\ie the wavelength of the polariton)
is large compared to $\vec k_\parallel$.
It should be noted that this is not necessarily as good an assumption as the time-scale separation,
because of the high confinement of the plasmon polaritons.

To the best of our knowledge, there is no model for the nonlocal effect in the intraband
(Drude-like) conductance that is both applicable beyond the interband threshold ($s>2$)
and conducive for the style of analytical calculations we present in this paper.
Therefore, we restrict ourselves to the effect of nonlocality on the interband case, which
is anyway potentially of greater interest, because of its resonant nature.

The origin of the nonlocal interband effect is the conservation of momentum, where we
assume a large graphene sheet and homogeneous optical illumination (spatially slowly
varying envelope):
A carrier from the lower branch of the dispersion relation with an initial wave vector
on the circle $|\vec q_\mathrm{in}| = q_0$ is not lifted to 
the same wave vector on the upper branch, but to a final wave vector on the circle that 
is shifted by the optical wave vector $\vec k_\parallel$:
$|\vec q_\mathrm{fin} - \vec k_\parallel| = q_0$.
As a result, the effective interband transition energy is no longer independent of the
exact wave number on the initial wave number circle, but offset by up to 
$\pm v_\Fermi \vec k_\parallel$.
Since all source states with the same energy form a circle in $\vec k$-space,
we find for the nonlocally corrected conductance:
\begin{align}
  \sigma_2(\omega, k_\parallel) = \frac{1}{\pi}
\int_{-\pi}^{0} \total \alpha \ \sigma_2(\omega + v_\Fermi k_\parallel \cos \alpha),
\end{align}
where the conductivity under the integral is the local conductivity and only the 
modulus $k_\parallel$ of the optical wave vector matters
because of the cylindrical symmetry of the Dirac cone.
This smearing effect carries through the entire derivation and does not interact
with the partial derivatives leading to the ponderomotive force.
Hence, we find for the interband ponderomotive force including nonlocal corrections:
%\begin{align}
%\vec F_2(\omega, k_\parallel) = \int_{-1}^1 %\total \alpha \ \frac{
%\vec F_2(\omega + v_\Fermi k_\parallel %\alpha)}{\sqrt{1 - \alpha^2}} .
%\end{align}
\begin{align}
\vec F_2(\omega, k_\parallel) =& \frac{1}{\pi} \int_{-1}^1 \total z \, \frac{
\vec F_2(\omega + v_\Fermi k_\parallel z)}{\sqrt{1 - z^2}},
\end{align}
with $z = \cos \alpha$.
We only study the seemingly more complex case of finite temperature, because at
$T=0$, there is a pole in the integration interval, which means that the zero
temperature case must be obtained as the limit $T\rightarrow0$ of the finite
temperature expression.
As before, we choose to separate the normalized frequency dependence from 
constants that clutter the notation by generalizing the inter-band contribution
$P_2(s)$ to the normalized response function $P(s)$:
\begin{align}
P_2(s,\phi)=&\frac{1}{\pi}\int_{-1}^{1}\total z \, \frac{P_2(s+\phi z)}{\sqrt{1-z^2}},
  \label{eqn:nonlocal_def_normalized}
\end{align}
where $\phi=k_\parallel /k_\Fermi$ is the normalized in-plane optical wave number
and therefore simultaneously parameterizes the optical phase velocity and angle
of incidence.
As far as we can tell, Eq.~\eqref{eqn:nonlocal_def_normalized} has no closed 
solution in terms of fundamental functions, but under the assumptions 
$s\approx2$ and $\phi \ll 1$ ($c \gg v_\Fermi$), we can approximate it (see 
Appendix~\ref{appx:nonlocal} for details):
\begin{subequations}
\begin{align}
  \nonumber
  P_2(s,\phi) 
  = & 2 I(s, \phi, -2),
  + I(s, \phi, 2 + 2\imag \theta) 
  \\
  & \quad 
  + I(s, \phi, 2 - 2\imag \theta) 
  \intertext{with}
  I(s, \phi, A) \approx & 
  \frac{(2s - A) \sgn(\Re\{s-A\}) }{2 s^2 \sqrt{(s - A)^2 - \phi^2}}
  - \frac{1}{2s^2},
\end{align}
\label{eqn:nonlocal_approx}
\end{subequations}
where $\sgn(\Re\{z\})=x/|x|$ for $z=x+iy$. We find that this approximate expression
matches a direct numerical integration
of Eq.~\eqref{eqn:nonlocal_def_normalized} very well within in the range 
$1.5 \leq s \leq 2.5$ and $\phi \leq 0.1$, which corresponds to an optical 
wavenumber ten times higher than in vacuum.

In Fig.~\ref{fig:nonlocal_approx}, we show Eq.~\eqref{eqn:nonlocal_approx} for 
two normalized temperatures and a number of normalized in-plane wave numbers.
We find that the effect of the interband nonlocality on the ponderomotive
force differs significantly from most nonlocal corrections found in plasmonics.
Usually, nonlocality leads to spectral shifts as well as additional damping and smearing of spectral 
features \eg due to Landau damping~\cite{Raza:2015}.
Instead, we find a splitting of the ponderomotive resonance in the regime
$\phi>\theta$ corresponding to $k_\parallel > k_B T /\hbar v_\Fermi$.
Below this threshold, the splitting exists in principle, but is hidden by 
thermal broadening.
%
%%%%%%%%%%%%%%%%%%%%%%%%%%%%%%%%%%%%%%%%%%%%%%%%%%%%%%%%%%%%%%%%%%%%%%%%%%%%%%%%
\section{Conclusion}

In summary, we derived an expression for the ponderomotive force arising from the
optical excitation of a polariton in a 2D material, focusing our analysis on the
case of graphene. Starting from a thermodynamic approach that relates the material
permittivity to the total free enthalpy, we obtained a general relation between
the ponderomotive force and the 2D sheet conductivity. By introducing into this
the appropriate expression for the interband graphene conductivity, we showed that the
resulting ponderomotive force exhibits a pole, which leads to its divergence at
zero temperature. At higher temperatures, this divergence is gradually smoothed,
but there is always an energy region around half the Fermi energy where the
interband contribution is larger than the corresponding Drude part. This can
play an important role when exploring second-order intrinsic nonlinearities in
polaritonic materials, as for example in the recent study of stimulated plasmon
polariton scattering~\cite{Wolff:2018}.
Finally, we explored the impact of nonlocal corrections to the interband part
assuming a spatially slowly varying optical envelope, and found a very 
characteristic type of resonance splitting.
Strikingly, this is very distinct from the simple broadening and
minuscule resonance shifts, which are the most common type of nonlocal 
correction in linear plasmonics~\cite{Raza:2015b,Christensen:2014,Christensen:2017}, 
and therefore a very clear signature for nonlocal response in plasmonics.

%%%%%%%%%%%%%%%%%%%%%%%%%%%%%%%%%%%%%%%%%%%%%%%%%%%%%%%%%%%%%%%%%%%%%%%%%%%%%%%%
\section*{Acknowledgments}
C.~W. acknowledges funding from a MULTIPLY fellowship under the Marie
Sk\l{}odowska-Curie COFUND Action (grant agreement No. 713694).
The Center for Nano Optics is financially supported by the University of
Southern Denmark (SDU 2020 funding).
C.~W. also acknowledges controversial yet fruitful discussions with Dr Wolff.
N.~A.~M. is a VILLUM Investigator supported by VILLUM Fonden (grant No. 16498).
The Center for Nanostructured Graphene is sponsored by the Danish National 
Research Foundation (Project No. DNRF103). 

%%%%%%%%%%%%%%%%%%%%%%%%%%%%%%%%%%%%%%%%%%%%%%%%%%%%%%%%%%%%%%%%%%%%%%%%%%%%%%%%
\begin{appendix}
\section{Analytic approximation to the nonlocal correction}
\label{appx:nonlocal}

\begin{widetext}
Here, we find an approximate solution to the integral
\begin{align}
  P_2(s,\phi) = & \frac{1}{\pi} \int_{-1}^{1} \total z \ \frac{1}{\sqrt{1-z^2}}
  \frac{1}{s + \phi z}
  \left[\frac{s + \phi z - 2}{(s + \phi z - 2)^2 + 4 \theta^2} 
  + \frac{1}{s + \phi z+ 2}
  \right]
  \\
  = & \frac{1}{2\pi} \int_{-1}^{1} \total z \ \frac{1}{\sqrt{1-z^2}} \cdot
  \frac{1}{s + \phi z} \left[
    \frac{1}{s + \phi z - 2 + 2\imag \theta} 
    + \frac{1}{s + \phi z - 2 - 2\imag \theta} 
    + \frac{2}{s + \phi z + 2}
  \right].
  \label{appx1:P2_partial_fraction}
\end{align}
This partial fraction decomposition reduces the problem to three integrals of 
the same form
\begin{align}
  I(s, \phi, A) = \frac{1}{2\pi} 
  \int_{-1}^{1} \total z \ \frac{1}{\sqrt{1-z^2}} \cdot
  \frac{1}{s + \phi z} \cdot \frac{1}{s + \phi z - A}.
\end{align}
Next, we restrict ourselves to the neighborhood of the inter-band resonance,
\ie $s\approx 2$ and we assume $\phi \ll 1$.
This means we may approximate $(s + \phi z)^{-1} \approx s^{-2} (s - \phi z)$:
\begin{align}
  I(s, \phi, A) 
  \approx & \frac{1}{2\pi s^2} \int_{-1}^{1} \total z \ \frac{1}{\sqrt{1-z^2}} 
  \cdot \frac{s - \phi z}{\phi z + s - A}
  = \frac{1}{2\pi s^2} \int_{-1}^{1} \total z \ \frac{1}{\sqrt{1-z^2}} 
  \left[
    \frac{2s - A}{\phi z + s - A}
    -1
  \right]
  \\
  = & \frac{-1}{2 s^2} + \frac{2s - A}{2 \pi \phi s^2}
  \int_{-1}^{1} \total z \ \frac{1}{(z + a) \sqrt{1-z^2} },
  \quad\quad \text{with} \quad a = \frac{s - A}{\phi}.
\end{align}
This is a standard integral and evaluates to $\pi / \sqrt{a^2 - 1}$.
This expression is ambiguous as to which branch of the root function is 
to be used.
This is solved by identifying the integration result at $\phi = 0$ with the 
local expression $P_2(s)$.
In this way, we arrive at Eq.~\eqref{eqn:nonlocal_approx}.

\end{widetext}

\end{appendix}

\bibliography{ref}

\end{document}